\documentstyle[prl,aps,psfig]{revtex}
\begin{document}
\draft
\title{The  Composite Fermion Hierarchy:
Condensed States of Composite Fermion Excitations?}
\author{P. Sitko$\sp{1,2}$, K.-S. Yi$\sp{3}$,
 and  J. J. Quinn$\sp{1,4}$}
\address{
$\sp{1}$University of Tennessee, Knoxville, Tennessee 37996, USA
}
\address{
$\sp{2}$Institute of Physics, Technical University of Wroc{\l}aw, 
Wyb. Wyspia\'nskiego 27, 50-370 Wroc{\l}aw, Poland
}
\address{ 
$\sp{3}$Pusan National University, Pusan 609-735, Korea
}
\address{
$\sp{4}$Oak Ridge National Laboratory, Oak Ridge, Tennessee, 37831, USA
}
\date{\today}
\maketitle

\begin{abstract}
A composite Fermion hierarchy theory is constructed
in a way related to the original Haldane picture 
by applying the composite Fermion (CF)
transformation to quasiparticles of Jain states. 
 It is shown that
the Jain theory
coincides with the Haldane hierarchy theory for principal
CF fillings. 
Within
the Fermi liquid approach for few electron systems on the sphere
a simple interpretation of many-quasiparticle spectra is given
and provides an explanation of failure of CF hierarchy picture 
when applied to the hierarchical $4/11$ state.
\end{abstract}

\narrowtext
The Haldane hierarchy \cite{Haldane,Prange} 
of fractional quantum Hall
states is obtained when quasiparticle excitations of  Laughlin \cite{Laughlin}
$\nu=1/m$ incompressible states undergo condensation into  new 
incompressible (daughter) states. 
Jain \cite{Jain} noted that certain hierarchy fractions appeared to be 
particularly stable, and he introduced the composite Fermion (CF)
picture which predicts incompressible quantum
liquid states at fillings $\nu=n(1+2pn)^{-1}$, where $p=1,2,3,...$, and
$n=\pm 1,\pm 2,\pm 3, ...$ .
By invoking electron-hole symmetry
 condensed CF states at filling factor $1-|\nu|$ can be
found. These two Jain sequences do not contain all odd denominator 
fractions which are present in the hierarchy theory.
The missing odd denominator fractions can be generated
by introducing an operator $L$ whose application to a Jain trial 
wavefunction increases the CF filling factor by unity \cite{Goldman}.
Because the application of $L$ to a FQH state  weakens it enormously,
Jain and Goldman suggest that to a first approximation
$L$ can be completely neglected. Instead of constructing trial
wavefunctions, Lopez and Fradkin \cite{Lopez} used the CF transformation
to construct a conventional many body description of the interactions,
both Coulomb and Chern-Simons gauge interactions,
among the fluctuations beyond the mean field. They considered
only states belonging to the principal Jain sequence,
so it is unclear how the hierarchy of odd denominator
fractions appears in their approach.

The object of the present paper is to demonstrate how the hierarchy
of CF states arises.
It involves applying the CF transformation to quasiparticles
(CF's in a partially filled shell) and assuming that mean field
theory adequately describes the resulting state.
The procedure is not at all obvious, since
it essentially divides the electrons into different classes.
Furthermore, the validity of mean field theory rests on the
assumption that the gap for creating new quasiparticles
is large compared to their interaction.
Direct comparison of the CF hierarchy predictions with numerical
results for small systems displays a number of cases
in which this assumption fails.

The state with fractional filling $\nu_0$ has an average
of $\nu_0^{-1}$ flux quanta per electron. Following Jain,
we attach to each electron an even number, $2p_0$, of flux quanta
oriented opposite to the applied magnetic field.
This gives an effective CF filling factor $\nu_0^*$ given by $(\nu_0^{*})^{-1}=
\nu_0^{-1}-2p_0$. If $\nu_0^*$ is equal to an integer $n_1$, an integral number
of CF levels (or shells) is filled, and the Jain sequence 
$\nu_0=n_1(1+2p_0n_1)^{-1}$ is obtained. Negative $\nu_0^*$ indicates
that the effective magnetic field $B^*$ seen by a CF is oriented
opposite to the applied field $B$.

If $\nu_0^*$  
is not  an integer it can be written as $\nu_0^*=n_1+\nu_1$,
where $\nu_1$ represents the fractional filling of the partially
filled CF shell.
We define particles in the partially
filled CF shell as quasiparticles of the incompressible CF state with 
filling $\nu_0^*=n_1$ \cite{SNYi}.
If excitations
between Landau levels are negligible (and closed shells give nothing but
constant shift in energy)
 the lowest energy sector 
is determined by the energy states of the partially filled shell
\cite{shellbook,Sitko}. 
The same prediction of
the incompressible states as for the analogous
partially filled electron shell are found when
the CF transformation
is applied to quasiparticles and $(\nu_1^*)^{-1}=\nu_1^{-1}-2p_1$.
If  $\nu_1^*$ is still not an integer we can repeat the process 
which leads us to the sequence of the following equations:
\begin{equation}
\nu_l^{-1}=2p_l+(n_{l+1}+\nu_{l+1})^{-1} .
\end{equation}
From Eq. (1), the original fractional electron filling $\nu_0$,
can be  represented in a form of a continued fraction.
If $n_l=1$ for every value of $l$, this fraction can be put in the form
equivalent to Haldane's \cite{Haldane} (the particle-hole symmetry
can be additionally included at every step).
The hierarchical incompressible states are predicted when $\nu_{l+1}=0$
at any step.
If $\nu_1=0$, the principal Jain sequence is obtained.

It is interesting to see how the Jain principal states
arise in the Haldane picture (Eq. (1) with $n_{l+1}=1$).
Let us write the Haldane sequence in the following way
\begin{equation}
\label{Jain1}
\nu_l^{-1}=2p_l+1 -(1+\nu_{l+1}^{-1})^{-1} .
\end{equation}
Hence, even if all $p_l=0$, the hierarchical fraction can be generated.
After repeating procedure (\ref{Jain1}) with all
$p_l=0$ (except $p_0$) till $\nu_{n+1}=0$ 
(then $\nu_{n}=1$, $\nu_{n-1}=2$, ...,
and $\nu_1=n$) we have for $\nu_0$:
{\large
\begin{equation}
\label{Jain2}
\begin{array}{c}
\frac{1}{2p_0+1-\frac{1}{2-\frac{1}{
2-\frac{1}{2-... -\frac{1}{2}}}}} 
=\frac{n}{2p_0 n+1}.
\end{array}
\end{equation}}Hence, the Jain fraction 
can be obtained within the sequence (\ref{Jain1})
and the Haldane hierarchy theory coincides with Jain's for the fraction 
(\ref{Jain2}) 
(the equivalence of different hierarchy schemes was considered in
\cite{Read,Greiter,Ginocchio}).
Setting $n_l=1$ in the sequence (1) does
not omit any fraction; simply the higher values of $n$ are obtained in
the more complicated way.

Let us consider
a spherical system containing  $n_l^{QE}$ quasielectrons
at the $l^{th}$ hierarchical level. If those particles
occupy one filled Landau shell and there are extra
particles in the partially filled second Landau level 
one finds for the lowest angular momentum shell
 $2S_{l+1}-1=n_l^{QE}-n_{l+1}^{QE}$ \cite{Haldane,Chen}.
A condensed state
of these quasiparticles is obtained when they fill an integral
number of shells at the next step of the hierarchy 
 ($n_{l+2}^{QE}=0$); 
then we have additionally
$2S_{l+1}=(2p_{l+1}+1)(n_{l+1}^{QE}-1)$.
Combining these two relations we get:
\begin{equation}
n_l^{QE}=(2p_{l+1}+2)(n_{l+1}^{QE}-1)
\end{equation}
in agreement with Haldane \cite{Haldane} (if quasiholes
are produced then:
$n_l^{QE}=2p_{l+1}(n_{l+1}^{QH}-1)$). Thus, we identify Haldane even numbers
as $2p_{l+1}+2$ for quasielectrons ($2p_{l+1}$ for quasiholes),
the odd number is given by $2p_0+1$. 
The construction of the Haldane hierarchy for quasielectrons
without applying the CF transformation  to them, i. e. $p_{l+1}=0$
(redefinition of the number of quasielectrons), 
leads to the Jain principal fraction (\ref{Jain2}), and the picture
becomes identical to the original Jain construction.
We would like to note that our result is not based on identifying
the quasiparticles as Bosons, rather our (equivalent to Haldane
results) were obtained treating quasiparticles as Fermions.

Let's look at a simple example for $N=8$ electrons on a sphere.
If the degeneracy of the lowest electron shell is $2S_0+1=19$,
then we can attach two flux quanta (oriented opposite to $B$)
to each electron to obtain $2S_0^*=2S_0-2(N-1)=4$.
For a finite system we substract $2p_0(N-1)$ flux quanta to obtain the 
``mean'' value of the flux seen be one CF. The lowest CF shell accomodate
$2S_0^*+1=5$ particles, the remaining three become quasielectrons
in the next shell. These CF excitations have angular momentum $l_{QE}=3$.
The allowed values of the total angular momentum $L=0\oplus 2\oplus 3\oplus 4
\oplus 6$ are obtained by addition of the angular momenta of three Fermions
in the $l_{QE}=3$ shell. 
In numerical calculations these 3QE states are seen as the lowest energy sector
of the eight electron spectrum for $2S_0+1=19$ (cf. Fig. 1A).
The energy states can be obtained within the Fermi liquid shell model.
We can ignore particles in the filled shell
and treat only quasiparticles introducing
two-body Fermi liquid quasiparticle interaction \cite{SNYi,Sitko}.
When we attach $2$ flux quanta to each quasiparticle we find new CFs with
 a new value
$2S_{QE}^*=2S_{QE}-2(N_{QE}-1)=2$. Thus the effective shell for these
quasielectrons
can accomodate $2S_{QE}^*+1=3$ particles, exactly the number we have.
Within the CF hierarchy, the filling fraction is determined by
 $2p_0=2$, $n_1=1$, and $\nu_1$ (for quasielectrons) is given by
$2p_1=2$, $n_2=1$, and $\nu_2=0$. From Eq. (1) we find
$\nu_1^{-1}=3$ and  $\nu_0^{-1}=11/4$.

This picture can be applied to any system consisting of a finite number of
electrons on a spherical surface with a magnetic monopole
of strenght $2S$ at its center. For example, for the $N=6$ and $N=8$ electron
systems the hierarchy of condensed states for values of $\nu$ falling
between $1$ and $2/7$ is presented in Table I and 
Table II. 
The states in these tables that are marked with stars 
are not states in the principal Jain sequence.
They are hierarchical states
which can be obtained by following Haldane's original suggestion
and treating the quasiparticle excitations in the same way as the original 
electrons in the partially filled electron shell were treated.
In the Jain-Goldman \cite{Goldman} terminology,
these states are obtained with application of the
operator $L$.
Of these states only 
$6/17$ and $6/19$ states 
for $N=6$ and 
$\nu=11/17$ and $\nu=9/31$ states for $N=8$
have numerically observed  the $L=0$ ground state (but the gaps
are very small).
It is worth noting that 
that at fillings of the form $\nu_0=n_0+\nu'$,
where $n_0$ is an integer
and  $\nu'$ is an incompressible fractional
filling, 
incompressible states should occur.

The main advantage of the composite fermion theory is that it not only
predicts the ground states for the specific values of the magnetic field
($S$) but it also predicts the lowest energy states for every value of $S$
\cite{Sitko}.
Furthermore, when the CF transformation
and the MF approximation are applied to $N$ Fermions of angular momentum
$S$, the resulting energy spectra are the same whether the Fermions
are electrons, CF quasielectrons or CF quasiholes.
Figure 1  gives the  energy spectra of three  quasielectrons
in the angular momentum shells of 
$S_{QP}=3.0;\;3.5;\;4.0$ (indicated as A, B, C in the figure).
Figure 2 presents the cases of four quasielectrons for
$S_{QP}=3.5;\;4.0;\;4.5$.
Figures 3 and 4 are analogous results for three and four quasihole 
states. The crosses in these figures are the 
results of exact numerical
diagonalization within the lowest shell (only the lowest energy sector
is plotted), the open circles are Fermi liquid
theory results \cite{Sitko}.
The CF mean field approximation makes the following
predictions for the lowest energy states:
For three Fermions of angular momentum $S=3$,
an $L=0$ ground state;
for $S=3.5$ a single quasiparticle in the
$L=1.5$ state; for $S=4.0$ two quasiparticles 
each with $l=2$ should give
degenerate states $L=1\oplus 3$.
For the four Fermion systems, the $S=3.5$ gives
two quasiparticles each with $l=1.5$ resulting
in degenerate states $L=0\oplus 2$; $S=4$ gives
a single quasiparticle with $L=2$;
and $S=4.5$ gives an $L=0$ ground state.
We observe that while CF predictions hold for quasiholes (with small
overlap of the $L=0$ $2QP$ 
state with the higher energy states in Fig. 4A)
they fail to describe low lying 
quasielectron states (with one exception of $L=0$
ground state in Fig. 2C). 

A simple interpretation of the many-quasielectron spectra
comes from noting that QE interaction $V_{QE-QE}^S(J)$ 
 oscillates
with $J$ (pair angular momentum) exactly 
out of phase with the QH interaction \cite{SNYi} (for the same angular
momentum shell $S$).
Thus, to some extent, the 
shifts in the QE energies due to $QE-QE$
interactions are opposite to those of
QH's. 
We observe, in fact, that states which are highest energy states for 
quasielectrons are the lowest energy states for quasiholes.
The $1/3$ quasihole state, which is the $2/7$ Jain
state, is clearly observed (Fig. 3A, Fig. 4C). 
Hence, we can not expect
the same for the $1/3$ state of quasielectrons ($4/11$ state)
and the results for 3QE (Fig. 1A) and 4QE (Fig. 2C) spectra clearly show that,
though there is an $L=0$ ground state for $N=12$.
The qualitative similarity of  3QE and 4QE spectra (Fig. 1 and Fig. 2)
to reversed 3QH and 4QH spectra (Fig. 3 and Fig. 4)
 for other values of $S_{QE}=S_{QH}$ confirms our conclusion.


The authors would like to thank Geoff Canright for pointing out
a reference and discussion.
This work was supported in part by Oak Ridge National Laboratory, managed by
Lockheed Martin Energy Research Corp. for the US Department of Energy under
contract No. DE-AC05-96OR22464.
K.S.Y. acknowledges support by
 the BSRI--96--2412 program of the Ministry of Education, Korea.
P.S. acknowledges  support by Committee for Scientific Research, Poland,
 grant PB 674/P03/96/10.

\onecolumn
\begin{table}
\caption{$N=6$}
\begin{tabular}{|c||c|c|c|c|c|c|c|c|c|c|c|c|c|c|c|}
$2S+1$ & 
6 & 
7&
8 & 
9&
10 &
11&  
12 & 
13 & 
14 &
15 & 
16 &
17 & 
18 &
19 & 
20  \\ \hline
$\nu$ &
1 &
--&
6/7&
--&
2/3 &
--&
2/5 &
--&
 6/17$^*$ &
--&
 1/3 &
--&
6/19$^*$& 
--&
2/7 \\  
\end{tabular}
\end{table} 
\begin{table}
\caption{$N=8$}
\begin{tabular}{|c||c|c|c|c|c|c|c|c|c|c|c|c|c|c|c|c|c|c|c|c|}
$2S+1$ &
8 &
9 & 
10 & 
11 & 
12 & 
13 & 
14 &
15 & 
16 & 
17 & 
18 & 
19 & 
20 & 
21 &
22 &
23 & 
24 & 
25 &
26 & 
27 \\ \hline 
$\nu$&
1 & 
--&
8/9 &
4/5 &
9/13$^*$ &
2/3 &
11/17$^*$&
--&
11/27$^*$&
2/5 &
9/23$^*$ & 
4/11$^*$ &
8/23$^*$ &
--&
1/3 &
--&
8/25$^*$& 
4/13$^*$&
9/31$^*$&
2/7\\
\end{tabular}
\end{table}

\narrowtext
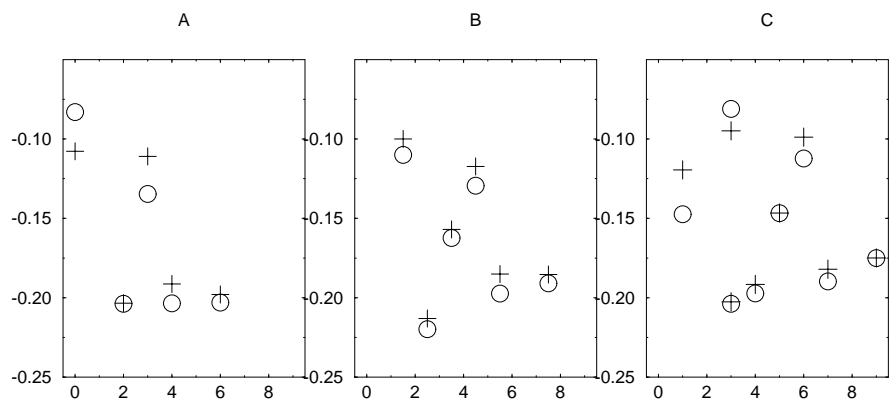
\begin{figure}
\caption{The spectra of three quasielectron systems
for $N=8,\;9,\;10$. The quasielectron interaction energy is given 
in units of $e^2/R$, 
where $R$ is the radius of the sphere. The $N=8$ case (Fig. 1A) 
is predicted to be  the $4/11$ state in the hierarchy theory. 
Circles represent Fermi liquid results and
crosses are exact numerical  energies.} 

\end{figure}

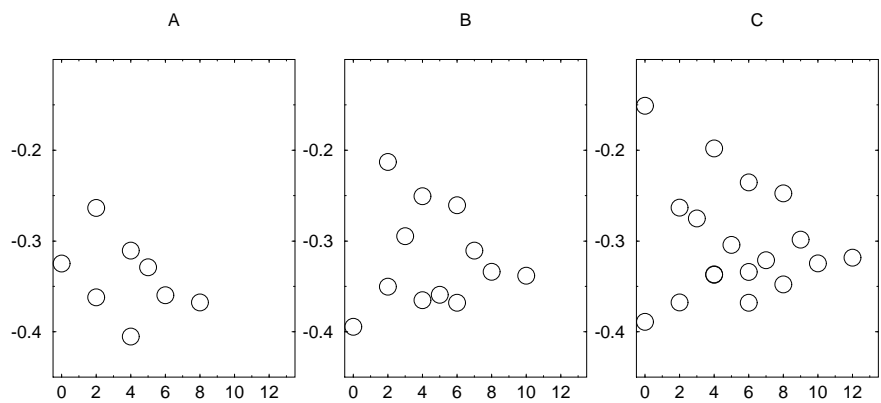
\begin{figure}
\caption{The Fermi liquid spectra of four quasielectron systems 
for
$N=10,\;11,\;12$. The $N=12$ case (Fig. 2C) corresponds to the $4/11$
hierarchical state.
The results are obtained using interaction parameters
found for systems of $N=8,\;9,\;10$, respectively.
The agreement with the numerical results 
is expected to be similar as that for 4QH spectra presented in Fig. 4.}
\end{figure}

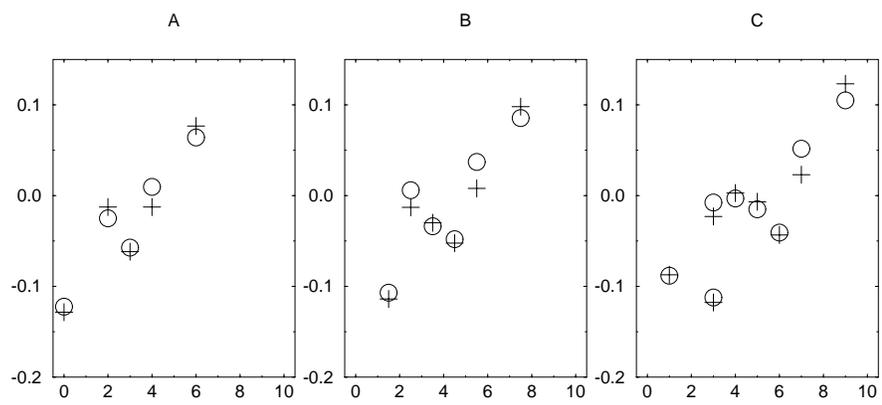
\begin{figure}
\caption{The spectra of 3QH for $S_{QH}=3.0;\;3.5;\;4.0$ ($N=4,\;5,\;6$).
The Fig. 3A ($S_{QH}=3.0$) shows the $2/7$ Jain state.}
\end{figure}

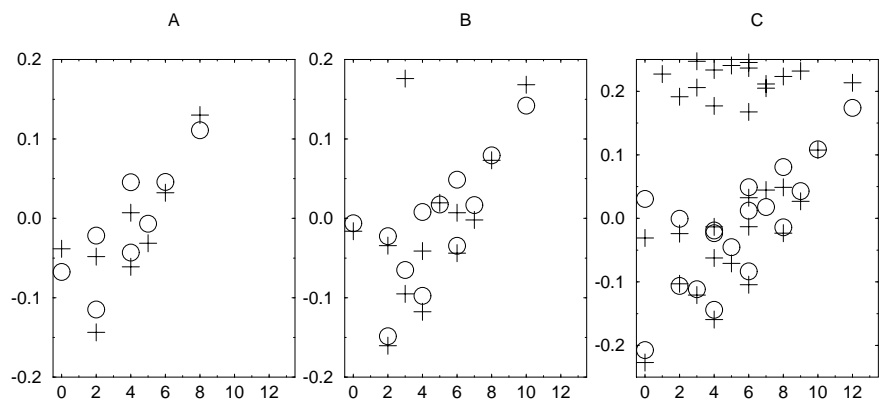
\begin{figure}
\caption{The spectra of 4QH for $S_{QH}=3.5;\;4.0;\;4.5$ ($N=4,\;5,\;6$).
The interaction parameters were obtained for systems of $N=6,\;7,\;8$,
respectively.
In Fig. 4C ($S_{QH}=4.5$) the spectrum of the $2/7$ state
is presented. The crosses not related to circles come from higher
energy states.}
\end{figure}

\end{document}